\documentclass[conference]{IEEEtran}
\usepackage{graphicx}
\usepackage{amssymb,amsmath}
\usepackage{tabu}
\usepackage{wrapfig}
\usepackage[nodisplayskipstretch]{setspace}
\usepackage{geometry}
\ifCLASSINFOpdf
\else
\fi
\begin{document}

\title{Skew-Induced Insertion Loss Deviation (SILD) and \text{FOM\_SILD}: Metrics for Quantifying P/N Skew Effects in High-Speed Channels}

  
 
\author{\IEEEauthorblockN{David Nozadze\IEEEauthorrefmark{1}\IEEEauthorrefmark{2},
Zurab Kiguradze\IEEEauthorrefmark{1},
Amendra Koul\IEEEauthorrefmark{1}, and
Mike Sapozhnikov\IEEEauthorrefmark{1}}
\IEEEauthorblockA{\IEEEauthorrefmark{1} Cisco Systems Inc., San Jose, CA, USA }
\IEEEauthorblockA{\IEEEauthorrefmark{2} dnozadze@cisco.com}}


\maketitle

\begin{abstract}
The rise of AI workloads and growing data center demands have driven the need for ultra-high-speed interconnects exceeding 200 Gb/s. As unit intervals (UI) shrink, even a few picoseconds of P/N skew can degrade serializer-deserializer (SerDes) performance. Traditional methods for quantifying skew fall short in capturing its impact. We introduce two new metrics: 1) Skew-Induced Insertion Loss Deviation (SILD) and 2) its complementary Figure of Merit (\text{FOM\_SILD}), analytically developed to assess P/N skew effects. Measured S-parameters confirm \text{FOM\_SILD} reciprocity, while simulations of 224G PAM4 SerDes show strong correlation with bit error rate (BER) trends. This approach offers a robust framework for analyzing skew in next-generation ultra-high-speed interconnects.
\end{abstract}
 \IEEEpeerreviewmaketitle

\section{Introduction}
\begin{figure}[!t]
	\includegraphics[width=3.5in]{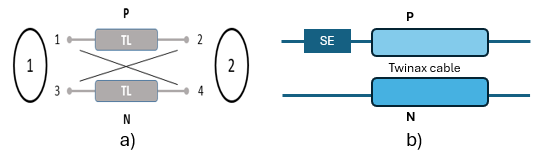}
	\caption{a) Schematic of Coupled Transmission Line: Ports labeled as 1 and 2 represent mixed-mode ports, while ports labeled as 1 through 4 correspond to single-ended ports. The P/N lines form a differential pair.		
		b) Schematic of Simulated Channels: P/N skew is introduced by placing single-ended (SE) delay lines in front of the twinax cable. The delay of the delay lines is adjusted to vary the skew in the channel from 0 to 3 ps.}
	\label{fig_1}
\end{figure}

As data rates increase, it becomes critical to evaluate all factors within a signal channel that can impact signal quality. At data rates exceeding 200 Gbps, one of the primary performance-limiting issues for high-speed Serializer/Deserializer (SerDes) links is P/N differential skew. P/N differential skew refers to the difference in arrival times between the two single-ended signals in a differential pair. This phenomenon is commonly attributed to asymmetry between the P and N traces of a differential pair (Fig.~\ref{fig_1} a). However, even in cases where the physical lengths of the P and N traces are perfectly matched, P/N skew can still arise due to asymmetries in the signal path.

One significant cause of P/N skew is the inhomogeneity of the materials used in high-speed interconnects, such as printed circuit boards (PCBs) and high-speed twinaxial cables. For PCBs, the variation in dielectric properties stems from the use of fiberglass and epoxy resin, which have different dielectric constants. Since the precise placement of fiberglass relative to the signal traces cannot be perfectly controlled during fabrication, random variations in the dielectric environment can introduce P/N skew. Similarly, in high-speed cables, variations in wire positioning or asymmetries in dielectric shielding along the cable length can lead to skew between the P and N signals.
Although numerous studies have attempted to measure and quantify P/N skew, a comprehensive methodology for predicting its impact on differential insertion loss (IL) in coupled channels remains incomplete \cite{ 2014_Tian_simp,2010_Miller_DC,2017_Nozadze_epeps, 2007_Loyer_CT,2017_Baek_DC,Nalla_2017_EMC,Nozadze_2017_EMC}.   Ref.~\cite{2021_Moon_spi} introduced a metric, termed EIPS, to analyze skew effects. However, this metric exhibits asymmetry when evaluating differential channels from left-to-right versus right-to-left, leading to non-unique values per channel and limiting its usefulness in assessing channel performance relative to skew.

In this paper, we introduce Skew-Induced Insertion Loss Deviation (SILD) and its complementary Figure of Merit (\text{FOM\_SILD}) to quantify the impact of P/N skew on high-speed differential channels. Using measured S-parameters, we validate their accuracy and demonstrate a strong correlation with bit error rate (BER) trends in high-speed SerDes simulations.
\section{P/N skew in Differential channels} 
\begin{figure}[t]
	\includegraphics[width=3.5in]{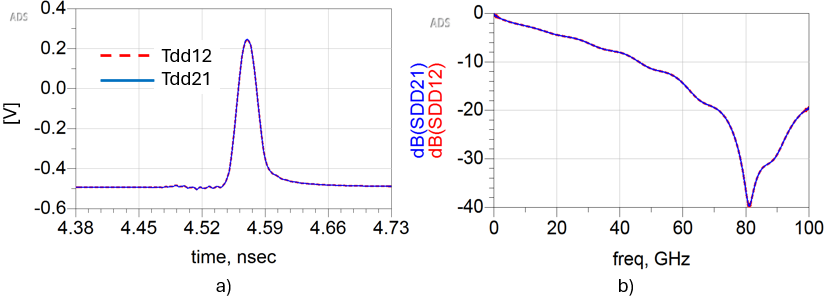}
	\caption{Differential-to-differential responses are shown for signal propagation: left-to-right in the blue line and right-to-left in the red line a) pulse responses, and b) through insertion losses.}
	\label{fig_4}
\end{figure}

In this section, impact of P/N skew on differential (IL) is studied using Transmission Line (TL) theory. To do so, coupled TL shown in Fig.~\ref{fig_1} a) is considered. The P/N phase skew at differential port 2 is defined as $t_{\rm{skew,2}}=t_{1,2}-t_{2,2}$ where  $t_{1,2}={\rm{phase}}(S_{sd21})/2\pi f$
and $t_{2,2}={\rm{phase}}(S_{sd41})/2\pi f$ are time delays corresponding to the propagation of the signal from  mixed-mode port 1 to single-ended ports 2 and 4, respectively. $S_{sd21}=1/\sqrt{2}(S_{21}-S_{23})$ and $S_{sd41}=1/\sqrt{2}(S_{43}-S_{41})$ are S-parameters
from mixed-mode port 1 to single-ended ports 2 and 4, respectively and $f$ is the frequency. Similarly, the P/N phase skew at differential port 1 would be
$t_{\rm{skew,1}}=t_{1,1}-t_{2,1}$ where  $t_{1,1}={\rm{phase}}(S_{sd12})/2\pi f$ and $t_{2,1}={\rm{phase}}(S_{sd32})/2\pi f$. Correspondingly, differential insertion losses would be
\begin{align}\label{diff_skew1}
	S_{dd21}=\frac{1}{2}\left(|S_{21}-S_{23}|e^{i 2\pi f t_{1,2}}+|S_{43}-S_{41}|e^{i 2\pi ft_{2,2}}\right)\,,
\end{align} 
and
\begin{align}\label{diff_skew2}
	S_{dd12}=\frac{1}{2}\left(|S_{12}-S_{14}|e^{i 2\pi f t_{1,1}}+|S_{34}-S_{32}|e^{i 2\pi ft_{2,1}}\right)\,.
\end{align} 
\begin{figure}
	\includegraphics[width=3.5in]{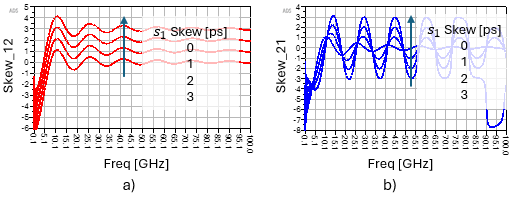}
	\caption{a) Simulated P/N phase skew signal propagating from right to left ($t_{\rm{skew,1}}$), and b) P/N phase skew signal propagating from left to right ($t_{\rm{skew,2}}$).}
	\label{fig_2}
\end{figure}
 \begin{figure}[b]
	\includegraphics[width=3.5in]{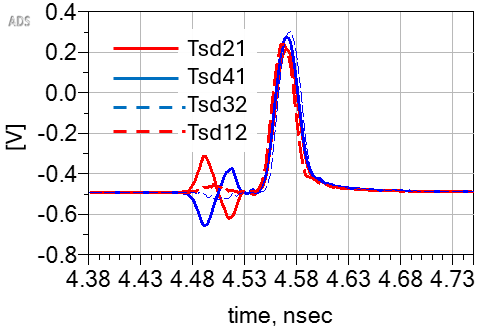}
	\caption{The solid line shows the differential-to-single-ended pulse response for left-to-right signal propagation, while the dashed line represents the response for right-to-left signal propagation.}
	\label{fig_3}
\end{figure}
In order to analyze skew behaviors from both sides, we use high-speed twinax S-parameters and add a single-ended delay line in front of one of the lines, e.g., on the P line. To vary the skew, we adjust the delay line's time delay $S_1$ from 0 to 3 ps (Fig.~\ref{fig_1} b). It has been observed that the P/N phase skew measured at differential port 1, $t_{\rm{skew,1}}$, is different from that measured at differential port 2, $t_{\rm{skew,2}}$ (Fig.~\ref{fig_2}). This indicates that the phase skews are not reciprocal. Therefore, P/N phase skews, $t_{\rm{skew,1}}$ and $t_{\rm{skew,2}}$, cannot be used as metrics to evaluate the skew's impact on the channel. Additionally, when examining the differential-to-single-ended mode pulses for left-to-right versus right-to-left propagation, the results are different (Fig.~\ref{fig_3}).
However, if we plot the differential insertion losses and the corresponding pulse responses for left-to-right versus right-to-left propagation, they are observed to be the same (Fig.~\ref{fig_4}). This implies that the S-parameters remain reciprocal. The reason for this is the fact that skew does not only impact the phase in coupled channels, but also affects the magnitudes of the differential-to-single-ended S-parameters, as reflected in the magnitudes in Eqs.~\ref{diff_skew1} and \ref{diff_skew2}. Thus, in addition to phase skew, we also have amplitude skew. It can be demonstrated that the phase shift, or so-called "amplitude skew," of $S_{21}$ versus $S_{23}$ ($S_{43}$ versus $S_{41}$) corresponds to $t_{\rm{skew,1}}$, and $t_{\rm{skew,2}}$ applies to $S_{12}$ versus $S_{14}$ ($S_{34}$ versus $S_{32}$). 
For more generic solutions see section IV. Consequently, the de-skewed magnitude of the differential insertion loss can be calculated by removing the extra phases in both the amplitude and phase components.
\begin{align}\label{de_skew1}
|S^0_{dd21}|=\frac{|S_{21}e^{-i 2\pi f t_{\rm{skew,1}}}-S_{23}|+|S_{43}-S_{41}e^{-i 2\pi f t_{\rm{skew,1}}}|}{2}\,
\end{align} 
and
\begin{align}\label{de_skew2}
|S^0_{dd12}|=\frac{|S_{12}e^{-i 2\pi f t_{\rm{skew,2}}}-S_{14}|+|S_{34}-S_{32}e^{-i 2\pi f t_{\rm{skew,2}}}|}{2}\,.
\end{align} 
This way, even though phase skews are not reciprocal, the de-skewed magnitude of differential losses for left-to-right versus right-to-left transmission will be the same. Thus, the impact of skew is fully removed from both magnitude and phase.
\section{P/N Skew metric} 
In this section, we will introduce new P/N skew metrics. Since we can de-skew the differential insertion loss magnitude correctly for either left-to-right or right-to-left transmission, we can calculate the so-called Skew-Induced Insertion Loss Deviation.
\begin{align}\label{SILD1_2}
	\mathrm{SILD}_{1(2)} = |\mathrm{S}_{dd12(21)}| - |\mathrm{S}^0_{dd12(21)}|\,.
\end{align} 
SILD is reciprocal and, therefore, is the same when calculated from left-to-right or right-to-left transmission. It measures the distortion of the differential insertion loss magnitude caused by P/N skew. The maximum absolute value of SILD within the signal bandwidth can be used as a metric. Additionally, we can derive another single-number skew metric by calculating the RMS value of SILD, similar to the figure of merit for insertion loss deviation defined in the IEEE 802.3 standard. Thus, the figure of merit for SILD will be:
\begin{align}\label{fom_sild}
	\mathrm{FOM\_SILD_{1(2)}}=\sqrt{\frac{1}{N}\sum_{i=1}^{N}\left({W}_i     
	     *\rm{SILD_{1(2)}}^2\right)}\,,
\end{align} 
where 
\begin{align}\label{PWF}
	W_i=sinc^2\left(\frac{f_i}{f_b}\right) \frac{1}{1+\left(f_i/f_r\right)^8}\frac{1}{1+\left(f_i/f_{t}\right)^4}\,.
\end{align}
The weight function $\rm{W}_i$ is defined in the IEEE 802.3 standard. Here, $f_b$ represents the signal rate, while $f_r$ and $f_{\rm{t}}$ denote the receiver 3 dB bandwidth and the 3 dB transmit filter bandwidth, respectively. The summation extends up to the maximum frequency for a given signal rate as defined in the IEEE 802.3 standard. 
$\rm{FOM\_SILD_{1(2)}}$ is reciprocal, unique number per channel and measures how much insertion loss gets distorted by the P/N skew.

\begin{figure}[t]
	\includegraphics[width=3.5in]{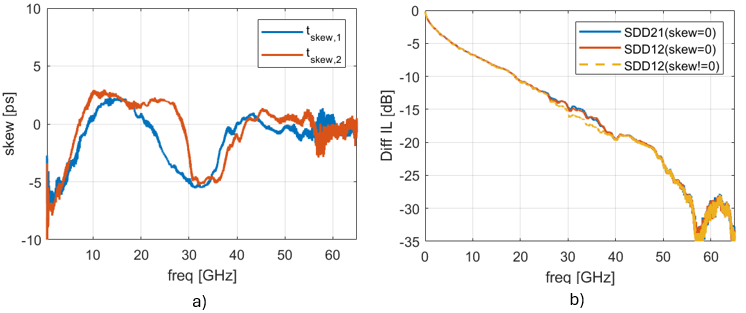}
	\caption{a) P/N phase skews of the measured twinax cable channel for right-to-left signal propagation ($t_{\rm{skew,1}}$) and left-to-right signal propagation ($t_{\rm{skew,2}}$).
		b) Differential insertion losses of the original twinax cable channel shown in the yellow line, and de-skewed differential insertion losses shown in the blue and orange lines.}
	\label{fig_5}
\end{figure}

To illustrate how our de-skew method works, we examine the measured S-parameters for channels designed for 224 Gbps PAM4.
Figure~\ref{fig_5} a) shows the P/N phase skews measured for left-to-right and right-to-left signal propagation. From Fig.~\ref{fig_5} a), it can be observed that the skew is highly non-linear as a function of frequency and not reciprocal, meaning that the measured phase P/N skews for left-to-right and right-to-left propagation are not the same. In Figure~\ref{fig_5} b), we see the magnitude of the de-skewed differential insertion losses compared to the original insertion losses. As shown, the de-skewed insertion losses significantly correct the insertion loss magnitude in the 25–35 GHz range (Fig.~\ref{fig_5} b), where skews are most pronounced (Fig.~\ref{fig_4} a).
Additionally, note that the corrected magnitudes of SDD21 and SDD12 are identical, demonstrating that the de-skew method preserves reciprocity. Correspondingly, the SILDs (Eq.~\ref{SILD1_2}) are also the same for left-to-right and right-to-left signal propagation.
\begin{figure}[b]
	\includegraphics[width=3.5in]{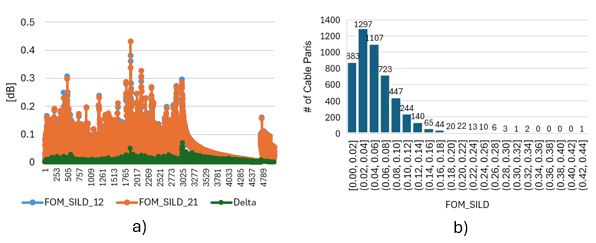}
	\caption{$\rm{FOM\_SILDs}$ for right-to-left signal propagation ($\rm{FOM\_SILD{12}}$) and left-to-right signal propagation ($\rm{FOM\_SILD{21}}$) measured across 5000 twinax cable channels. a) Delta represents the difference between ($\rm{FOM\_SILD{12}}$) and ($\rm{FOM\_SILD{21}}$). b) The distribution of  $\rm{FOM\_SILDs}$ for the same channels.}
	\label{fig_6}
\end{figure}
Next, we calculate $\rm{SILD_{1(2)}}$  and  $\rm{FOM\_SILD_{1(2)}}$ for over 5000 measured channels designed for 224 Gbps PAM4. These channels consist of PCBs and high-speed twinax cables.
Figure~\ref{fig_6} a) shows the $\rm{FOM\_SILD}$s values for both directions: right-to-left and left-to-right signal propagations. It can be observed that these values range from 0 to approximately 0.45 dB. In the same figure, the delta (the difference between $\rm{FOM\_SILD}$s values calculated for left-to-right and right-to-left signal propagations) is also plotted. As observed, the difference is very small, with only $0.34\%$ of the values exceeding 0.025 dB. Such differences can be attributed to measurement quality. This confirms that the metric is reciprocal and nearly identical for left-to-right and right-to-left signal propagation. Figure~\ref{fig_6} b) shows the distribution of $\rm{FOM\_SILD}$ values, where the majority are below 0.1 dB.

\begin{figure}[t]
	\includegraphics[width=3.5in]{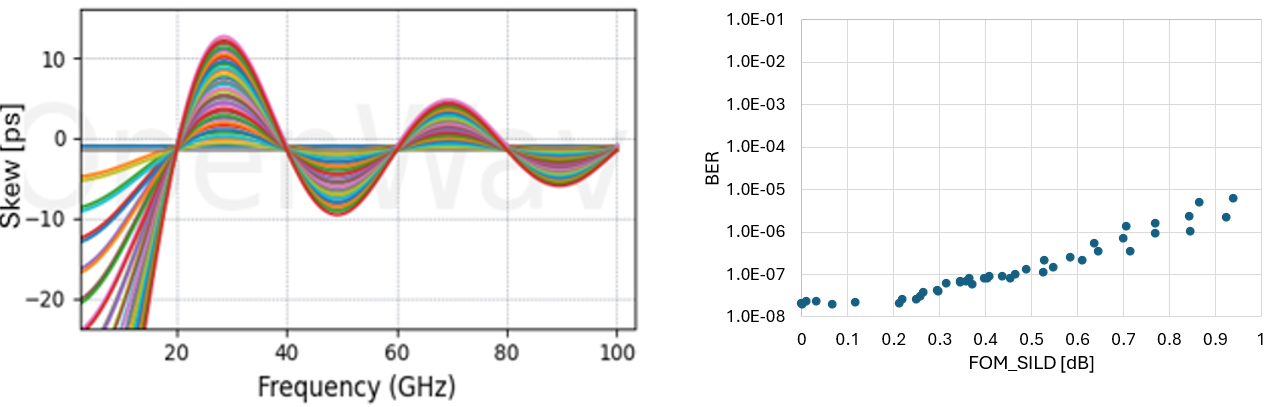}
	\caption{a) P/N skew profiles of simulated channels: flat, frequency-independent skew varies from 1 to 2 ps, while the oscillatory, frequency-dependent skew exhibits profile peaks ranging from 0 to 8 ps.
		b) Simulated 224 Gbps PAM4 BER versus $\rm{FOM\_SILDs}$ for channels with various skews from (a).}
	\label{fig_7}
\end{figure}

In addition, we validate through simulations that the  $\rm{FOM\_SILD}$ metric correlates with the BER performance of a 224 Gbps SerDes model. Channels are created with various skew levels and two skew profiles: a flat, frequency-independent profile and a damped oscillatory skew profile as a function of frequency (Fig.~\ref{fig_7} a). The details on how skew profiles are created is beyond the scope of this paper and will be provided in a forthcoming companion paper. Figure~\ref{fig_7} b) shows the relationship between BER and $\rm{FOM\_SILD}$. It can be observed that when $\rm{FOM\_SILD}$ is less than around 0.3 dB, the BER is not significantly impacted by skew. BER exponentially increases as $\rm{FOM\_SILD}$ goes up when $\rm{FOM\_SILD}$ is above 0.3 dB (see trend dashed line in Fig.~\ref{fig_7} b).
\vspace{7mm}

\section{De-skewing S-Parameters}
\begin{figure}[t]
	\includegraphics[width=3.5in]{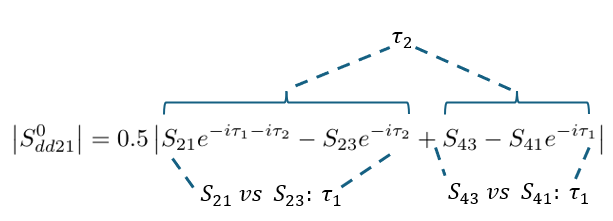}
	\caption{Schematic of S-parameters De-skewing.}
	\label{fig_10}
\end{figure}

In this section we will introduce method to de-skew S-parameters. Let $\tau_1$ be 
phase delay difference between THRU and FEXT (e.g. $S_{21}$ and $S_{23}$), and $\tau_2$ between differential and single ended S-parameters (e.g. $S_{sd21}$ and $S_{sd41}$) (see Fig.~\ref{fig_10}). Then, we can de-skew S-parameters as follows
\begin{equation}
	\begin{array}{c}
		S^0_{21} = S_{21}e^{-i\tau_1 - i\tau_2}, \quad S^0_{23} = S_{23}e^{-i\tau_2}, \quad S^0_{32} = S_{32}e^{-i\tau_2} \\
		S^0_{12} = S_{12}e^{-i\tau_1 - i\tau_2}, \quad  S^0_{14} = S_{14}e^{-i\tau_1}, \quad  S^0_{41} = S_{41}e^{-i\tau_1}.
	\end{array}\label{new_s_par}
\end{equation}

Based on the above notations (\ref{new_s_par}), the updated skew can be calculated as
\begin{equation}\label{new_skew}
	\begin{array}{c}
		t^0_{skew,2} = \frac{-i}{2\pi f} \left(\tau_2 + \tau_3 - \tau_4 \right), \\  t^0_{skew,1} = \frac{-i}{2\pi f} \left(\tau_1 + \tau_5 - \tau_6\right),
	\end{array}
\end{equation}
where
\[
\begin{array}{c}
	\tau_3 = phase\left(S_{21}e^{-i\tau_1} - S_{23}\right), \\
	\tau_4 = phase\left(S_{43} - S_{41}e^{-i\tau_1}\right), \\
	\tau_5 = phase\left(S_{12}e^{-i\tau_2} - S_{14}\right), \\
	\tau_6 = phase\left(S_{34} - S_{32}e^{-i\tau_2}\right).
\end{array}
\]

Assuming $t^0_{skew,1} = t^0_{skew,2} = 0$ in (\ref{new_skew}), we get 
\begin{equation}
	\begin{array}{c}
		\tau_2 + arctan\left(\frac{\frac{Im(z_1)}{Re(z_1)}-\frac{Im(z_2)}{Re(z_2)}}{1 + \frac{Im(z_1)}{Re(z_1)}\frac{Im(z_2)}{Re(z_2)}}\right) = 0, \\
		\tau_1 + arctan\left(\frac{\frac{Im(z_3)}{Re(z_3)}-\frac{Im(z_4)}{Re(z_4)}}{1 + \frac{Im(z_3)}{Re(z_3)}\frac{Im(z_4)}{Re(z_4)}}\right) = 0,
	\end{array}\label{system}
\end{equation}
where
\[
\begin{array}{c}
	z_1 = S_{21}e^{-i\tau_1} - S_{23}, \quad
	z_2 = S_{43} - S_{41}e^{-i\tau_1}, \\
	z_3 = S_{12}e^{-i\tau_2} - S_{14}, \quad
	z_4 = S_{34} - S_{32}e^{-i\tau_2}.
\end{array}
\]

\begin{figure}[t]
	\includegraphics[width=3.5in]{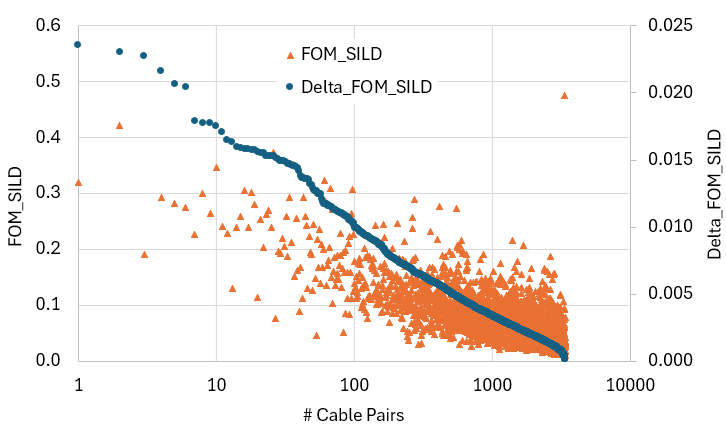}
	\caption{Numerically and analytically calculated $\rm{FOM\_SILD}$ for over 4000 measured cabled channels S-parameters.}
	\label{fig_9}
\end{figure}

De-skewed differential insertion losses will be
\[
\left|S^0_{dd21}\right| = 0.5\left|S_{21}e^{-i\tau_1 - i\tau_2} - S_{23}e^{-i\tau_2} + S_{43} - S_{41}e^{-i\tau_1}\right| =
\]
\[
= 0.5\left|S_{12}e^{-i\tau_1 - i\tau_2} - S_{14}e^{-i\tau_1} + S_{34} - S_{32}e^{-i\tau_2}\right| = \left|S^0_{dd12}\right|.
\]
Phase differences $\left(\tau_1, \, \tau_2\right)$, can be found using numerical methods for solution nonlinear system (\ref{system}) or analytically can be approximated as (\ref{diff_skew1}) and (\ref{de_skew2}) given in section II.
For measured  over 4000 cabled 112/224G PAM4 channels, we calculated both using numerical method and using analytical method (section II) $\rm{FOM\_SILD}$  and compare them. As shown in Fig. $\ref{fig_9}$ difference between accurate numerical method and approximated analytical method is very small. Difference is less than 0.015 dB for $\rm{FOM\_SILD}$ less than 0.1 dB and max is 0.023 dB for large $\rm{FOM\_SILD}$ of 0.32dB.

\vspace{7mm}

\vspace{7mm}

\section{Measurements} 

\begin{figure}[t]
	\includegraphics[width=3.5in]{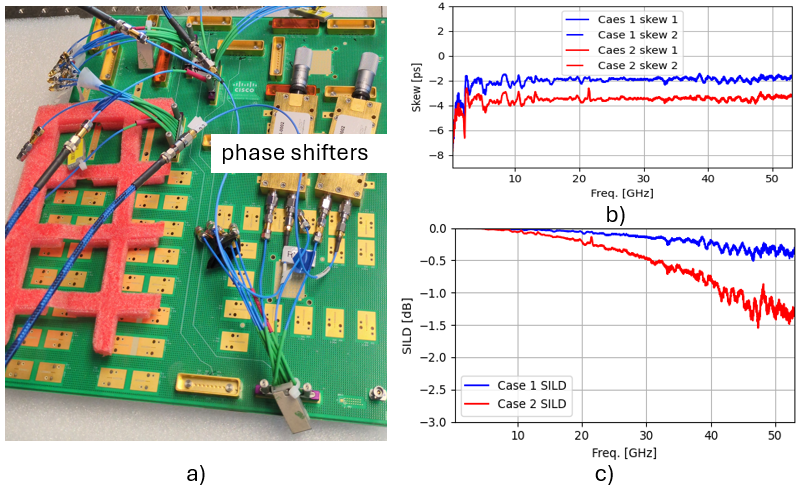}
	\caption{a) Measurement setup to create frequency-independent skew in the channel using phase shifters. b) Measured skew as a function of frequency for two skew values. c) SILD as a function of frequency for the two cases in (b).  Note that, due to the symmetry of SILD from left to right and right to left, only one side is shown in the plot.}
	\label{fig_11}
\end{figure}

\begin{figure}[t]
	\includegraphics[width=3.5in]{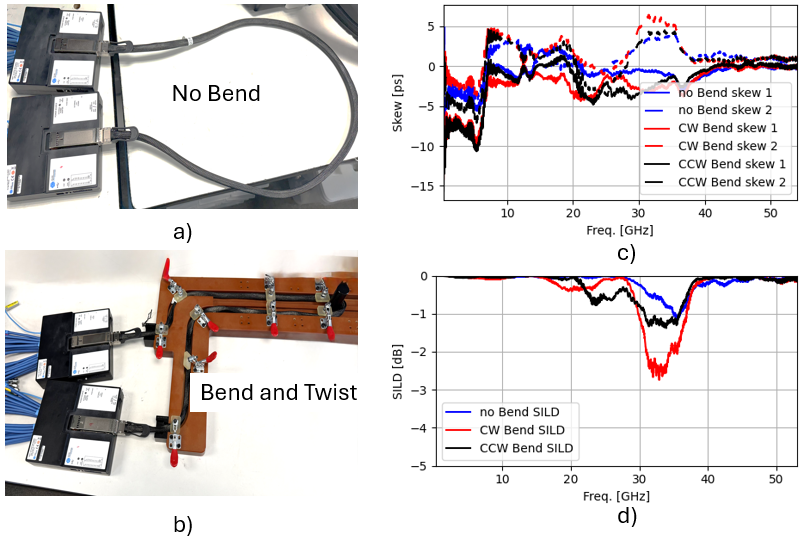}
	\caption{ a) Measurement setup for a 1.5 m DAC cable with no bend and no twist. b) Measurement setup for a 1.5 m DAC cable with bend and twist.  c) Measured skew as a function of frequency for the following cases: no bend/no twist, 360° clockwise twist (CW), 360° counterclockwise twist (CCW), and bend. d) SILD as a function of frequency for the two cases described in (b). Note that, due to the symmetry of SILD from left to right and right to left, only one side is shown in the plot.}
	\label{fig_12}
\end{figure}

In this section, we present BER measurements performed using a 224 Gbps SerDes IP. P/N skew in the channel was varied using a phase shifter to create frequency-independent skew (Fig.~\ref{fig_11}), and twinax cabled channels, which exhibit frequency-dependent skew (Fig.~\ref{fig_12}).

\vspace{7mm}

\vspace{7mm}

\begin{figure}[t]
	\includegraphics[width=3.5in]{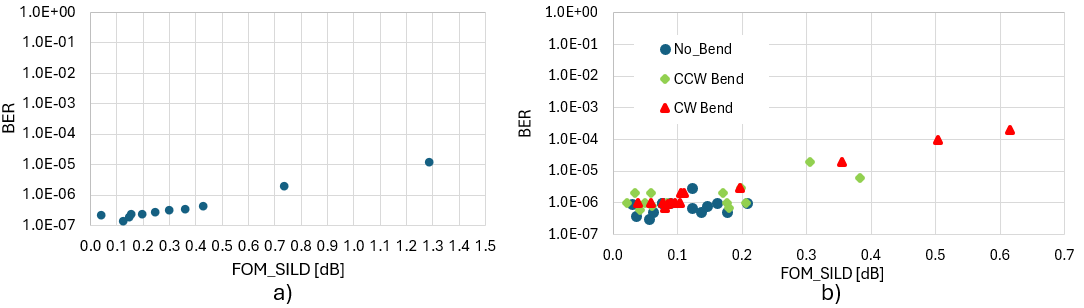}
	\caption{a) Measured BER  using a 224 Gbps SerDes IP vs $\rm{FOM\_SILD}$ for various frequency-independent P/N skews created in the channel using phase shifters.
		b) Measured BER using a 224 Gbps SerDes IP vs $\rm{FOM\_SILD}$for various frequency-dependent P/N skews in twinax cables, including cases with no bend/no twist, and with bends incorporating 360° clockwise (CW) and 360° counterclockwise (CCW) twists.}
	\label{fig_13}
\end{figure}
\vspace{7mm}

\subsection{Frequency independent skew} 

In this subsection, we present BER measurements conducted using a 224 Gbps SerDes IP, with the channel's P/N skew systematically varied through phase shifters to induce frequency-independent skew (Fig.~\ref{fig_11}). The skew was various distinct level, and the resulting BER was measured.

The measured BER as a function of $\rm{FOM\_SILD}$ is depicted in (Fig.~\ref{fig_13} a). Notably, when $\rm{FOM\_SILD}$ is below 0.3 dB, the BER remains relatively stable, indicating minimal impact on signal integrity. However, as $\rm{FOM\_SILD}$ exceeds 0.3 dB,  BER begins to increase noticeably.

\subsection{Frequency dependent skew}

In this subsection, we present BER measurements performed using a 
224 Gbps SerDes IP transmitted over twinax cable channels exhibiting frequency‑dependent skew
(Fig.~\ref{fig_12}). A 1.5 m DAC twinax cable was tested under three conditions:  1) no bend/no twist, 
2) 360° clockwise twist and bend  3) 360° counterclockwise twist and bend. BER plotted against the $\rm{FOM\_SILD}$ (Fig.~\ref{fig_13} b). Results indicate that for  $\rm{FOM\_SILD}$ values below 0.3 dB, BER remains largely unchanged, whereas a significant increase in BER is observed once  $\rm{FOM\_SILD}$ exceeds 0.3 dB. Cables subjected to bending and twisting exhibits higher  $\rm{FOM\_SILD}$ values and correspondingly degraded BER performance.

\section{Conclusions}
We demonstrated that P/N skew not only impacts the phase of differential-to-single-ended signal propagation but also affects its magnitude. We introduced new metrics: 1) SILD and
2) $\rm{FOM\_SILD}$ which are reciprocal. Through simulations of 224 Gbps PAM4 models and measurements using SerDes IP, we demonstrated that $\rm{FOM\_SILD}$ correlates with BER performance. Additionally, we presented measured statistics for over 5000 cabled channels and observed that $\rm{FOM\_SILD}$ is reciprocal, with the majority of cables having $\rm{FOM\_SILD}$ below approximately 0.1 dB.
One could define: 1) a maximum allowable target for SILD within the signal bandwidth and 2) a maximum value for $\rm{FOM\_SILD}$ as a specification to qualify channels.



\bibliographystyle{IEEEtran}
\bibliography{b}
 




\end{document}